\documentclass[twocolumn,twoside,slac_two]{revtex4}
\usepackage{graphicx}
\usepackage{fancyhdr}
\begin{document}

\title{Scalable Database Access Technologies \\ for ATLAS Distributed Computing}


\author{A. Vaniachine, for the ATLAS Collaboration}
\affiliation{ANL, Argonne, IL 60439, USA}

\begin{abstract}
ATLAS event data processing requires access to non-event data (detector conditions, calibrations, etc.) stored in relational 
databases. The database-resident data are crucial for the event data reconstruction processing steps and often required for 
user analysis. 
A main focus of ATLAS database operations is on the worldwide distribution of the Conditions DB data, which are necessary 
for every ATLAS data processing job. Since Conditions DB access is critical for operations with real data, we have developed the 
system 
where a different technology can be used as a redundant backup. Redundant database operations infrastructure fully satisfies the 
requirements of ATLAS reprocessing, which has been proven on a scale of one billion database queries during two reprocessing 
campaigns of 
0.5~PB of single-beam and cosmics data on the Grid. To collect experience and provide input for a best choice of technologies, 
several promising options for efficient database access in user analysis were evaluated successfully. We present ATLAS 
experience 
with scalable database access technologies and describe our approach for prevention of database access bottlenecks in a Grid 
computing environment.
\end{abstract}

\maketitle

\thispagestyle{fancy}
%\fancypagestyle{plain}{%
%\fancyhf{} % clear all header and footer fields
%\fancyhead[R]{ANL-HEP-CP-09-085 \\
%\begin{flushleft}
%To appear in the {\it Proceedings of the DPF-2009 Conference, Detroit, MI, July 27-31, 2009}
%\end{flushleft}
%} 
%\renewcommand{\headrulewidth}{0pt}
%\renewcommand{\footrulewidth}{0pt}}


%%%%%%%%%%%%%%%%%%%%%%%%%%%%%%%%%%
\section{Introduction}

A starting point for any ATLAS physics analysis is data reconstruction. ATLAS event data reconstruction requires access to 
non-event data (detector conditions, calibrations, etc.) stored in relational databases. These database-resident data are 
crucial for the event data reconstruction steps and often required for user analysis. Because Conditions DB access is critical for operations with real data, we have developed the system where a different technology can be used as a redundant backup. 

A main focus of ATLAS database operations is on the worldwide distribution of the Conditions DB data, which are necessary for every 
ATLAS data reconstruction job. To support bulk data reconstruction operations of petabytes of ATLAS raw events, the technologies 
selected for database access in data reconstruction must be scalable. Since our Conditions DB mirrors the complexity of the ATLAS 
detector~\cite{1}, the deployment of a redundant infrastructure for Conditions DB access is a non-trivial task.

%%%%%%%%%%%%%%%%%%%%%%%%%%%%%%%%%%
\section{Managing Complexity}

\begin{figure}[h]
\centering
\includegraphics[width=70mm]{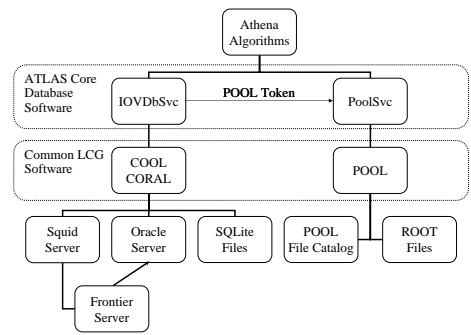}
\caption{Software for transparent access to several Conditions DB implementation technologies. Software for access to database-resident information is called CORAL, software for access to ROOT files is called POOL.} \label{db}
\end{figure}

Driven by the complexity of the ATLAS detector, the Conditions DB organization and access is complex (Figure~\ref{db}). 
To manage this complexity, ATLAS adopted a Conditions DB technology called COOL~\cite{2}.
COOL was designed as a common technology for experiments at the Large Hadron Collider (LHC). The LHC Computing Grid (LCG) project
developed COOL---Conditions Of Objects for LCG---as a subproject of an LCG project on data persistency called POOL---Pool Of 
persistent Objects for LHC~\cite{POOL}. The main technology for POOL data storage is ROOT~\cite{ROOT}.

In COOL the conditions are characterized by 
the interval-of-validity metadata and an optional version tag. ATLAS Conditions DB contains both database-resident information and 
external data in separate files that are referenced by the database-resident data. These files are in a POOL/ROOT format. ATLAS database-resident information exists in its entirety in Oracle but can be distributed in smaller ``slices'' of data using SQLite---a file-based technology.

\begin{figure*}[t]
\centering
\includegraphics[width=150mm]{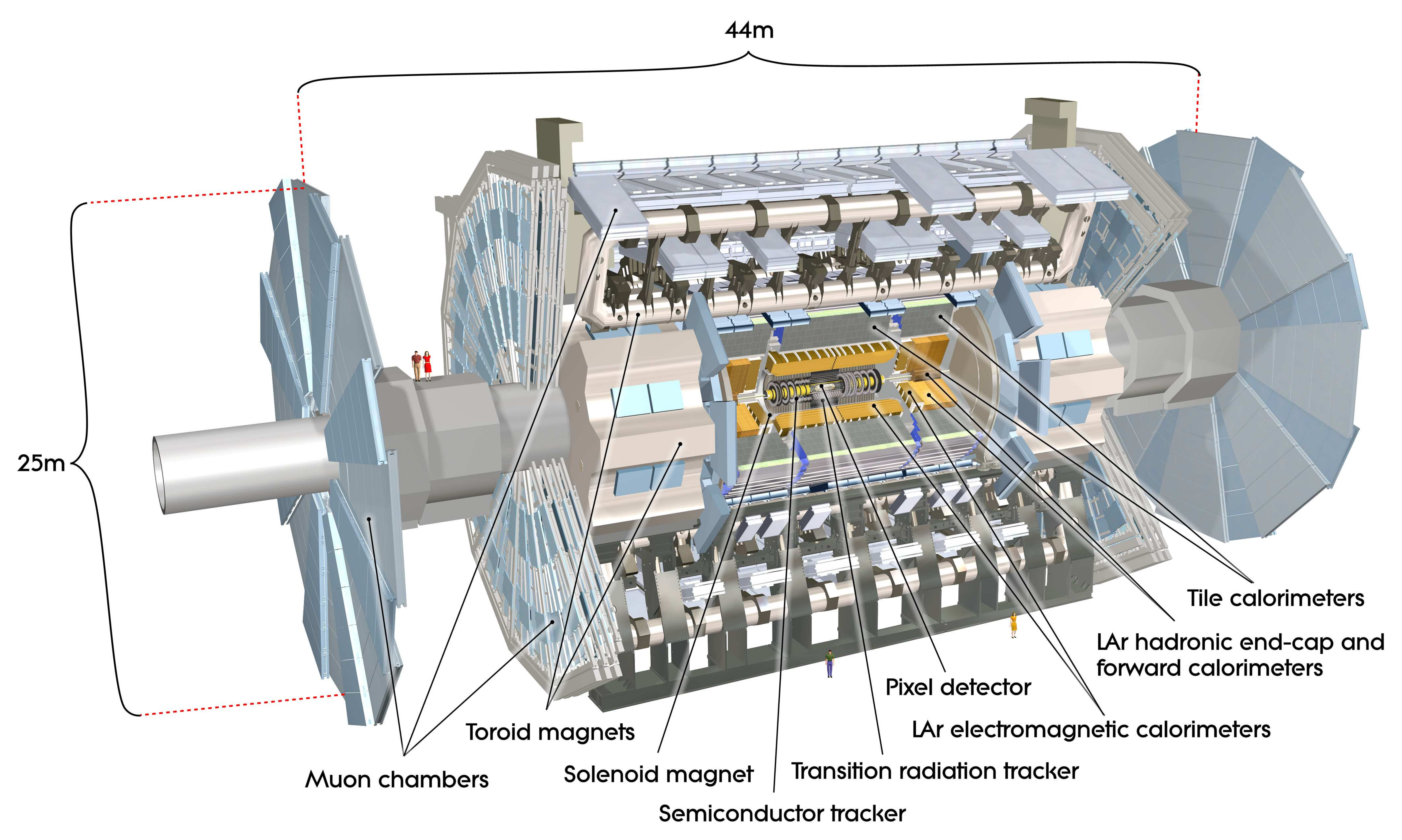}
\caption{Subdetectors of the ATLAS detector.} \label{atlas}
\end{figure*}

The complexity of the Conditions DB organization is reflected in database access statistics by data reconstruction jobs. 
These jobs access a slice of Conditions DB data organized in sixteen database schemas: two global schemas (online and offline) plus one or two schemas per each subdetector (Figure~\ref{atlas}). Jobs access 747 tables, which are grouped in 122 ``folders'' plus some system tables. There are 35 distinct database-resident data types ranging from 32 bit to 16 MB in size and referencing 64 external POOL files. To process a 2~GB file with 1000 raw events a typical reconstruction job makes $\sim$2000 queries reading  $\sim$40 MB of database-resident data, with some jobs read tens of MB extra. In addition, about the same volume of data is read from the external POOL files.

%%%%%%%%%%%%%%%%%%%%%%%%%%%%%%%%%%
\section{Data Reconstruction}

\begin{figure}[h]
\centering
\includegraphics[width=85mm]{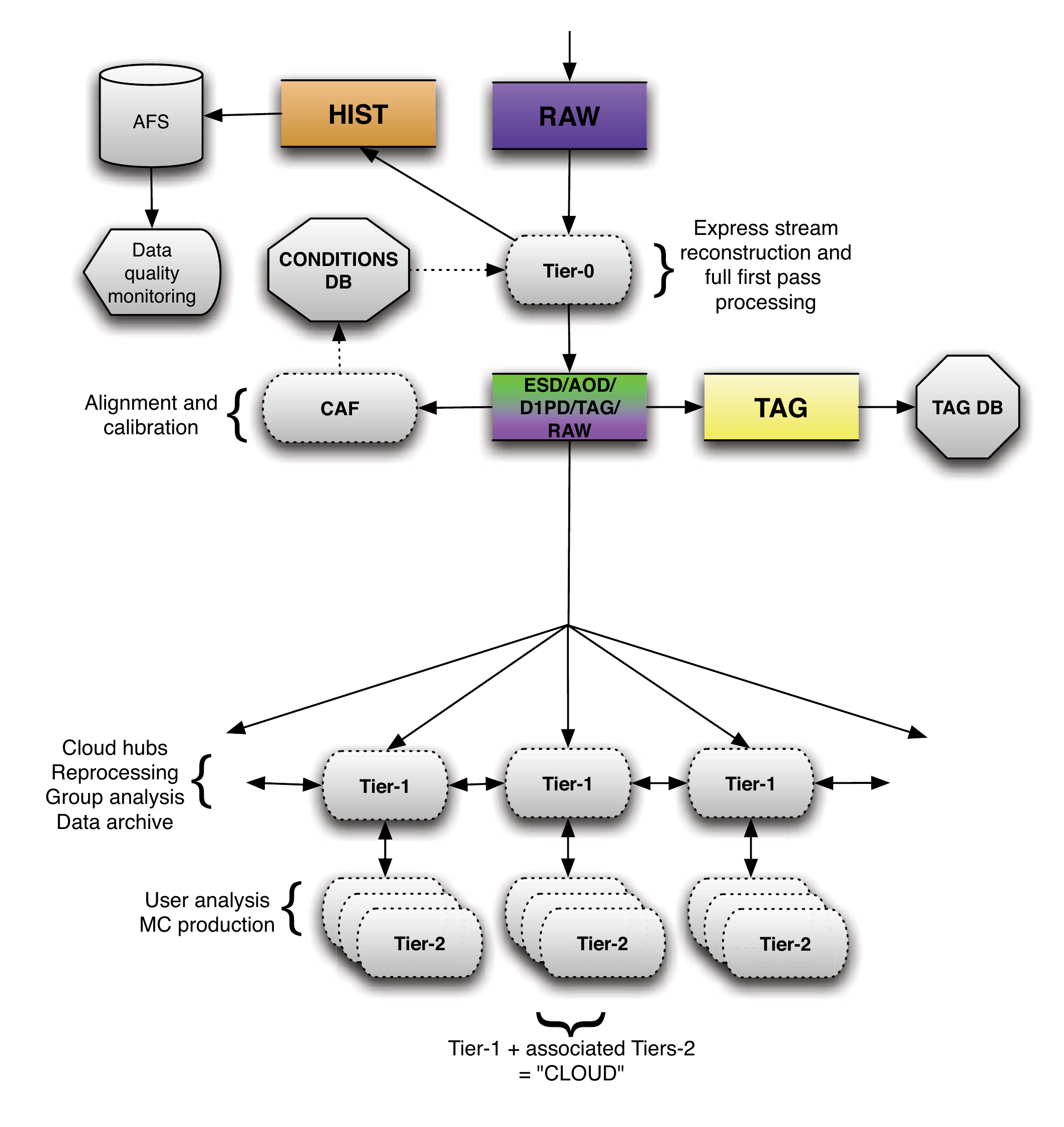}
\caption{Simplified flow of data from the detector (Fig.~\ref{atlas}) used in reconstruction at CERN and Tier-1 sites.} \label{flow}
\end{figure}

Data reconstruction is a starting point for any \linebreak[4] ATLAS data analysis. Figure~\ref{flow} shows simplified flow of raw events and conditions data in reconstruction.

%%%%%%%%%%%%%%%%%%%%%%%%%%%%%%%%%%
\subsection{First-pass processing at CERN}

Scalable access to Conditions DB is critical for data reconstruction at CERN using alignment and calibration constants produced within 24 hours---the ``first-pass'' processing. Two solutions assure scalability:
\begin{itemize}\addtolength{\itemsep}{-0.5\baselineskip}
  \item replicated AFS volume for POOL files,
  \item throttling of job submission at Tier-0.
\end{itemize}
The physics discovery potential of the Tier-0 processing results is limited because the reconstruction at CERN is conservative in scope and uses calibration and alignment constants that will need to be modified as analysis of the data proceeds. As our knowledge of the detector improves, it is necessary to rerun the reconstruction---the ``reprocessing.'' 
The reprocessing uses enhanced software and revised conditions for improved reconstruction quality. 
Since the Tier-0 is generally fully occupied with first-pass 
reconstruction, the reprocessing uses the shared computing resources, which are distributed worldwide---the Grid.

%%%%%%%%%%%%%%%%%%%%%%%%%%%%%%%%%%
\subsection{Reprocessing on the Grid}

\begin{figure*}[t]
\centering
\includegraphics[width=170mm]{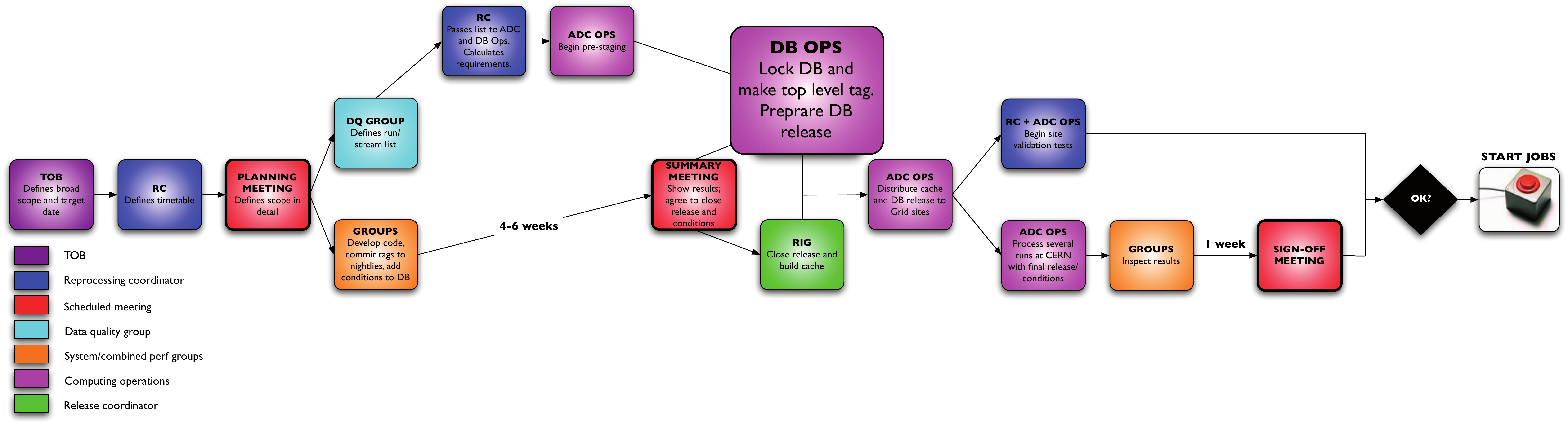}
\caption{Database Release build is on a critical path in ATLAS reprocessing workflow.} \label{workflow}
\end{figure*}

ATLAS uses three Grids (each with a different interface) split in ten ``clouds''. Each cloud consists of a large computing center 
with tape data storage (Tier\mbox{-}1 site) and associated 5--6 smaller computing centers (Tier-2 sites). There are also Tier-3 
sites---these are physicist's own computing facilities at the university or the department.

Reprocessing improves the particle identification and measurements over the first-pass
processing at CERN, since the reprocessing uses enhanced software and revised conditions.
Figure~\ref{workflow} shows reprocessing workflow that includes build of software and database releases.
To make sure that the results are of the highest quality obtainable, the full reprocessing campaigns on large fractions of the total 
data sample require months of preparation---these are the data that will be used in conferences and publications. As a result, most 
of the time in full reprocessing campaigns is occupied with validation of software and database releases, not actual running. 

To give faster feedback to subdetector groups we are doing reprocessing of smaller amounts of data, much quicker, to allow small 
modifications in software and conditions to be applied to previously processed data or as a contingency in case the 
Tier-0 ends up with a backlog of work. This is called ``fast'' reprocessing. 
It is also possible to do reprocessing not of the raw data but of the reconstructed data made during the last reprocessing campaign. This is called ESD reprocessing. The fast and ESD reprocessing are also performed on the Grid, 
in exactly the same way as ``full'' reprocessing.

\section{Database Access on the Grid}
\subsection{Database Release}

\begin{figure}[h]
\centering
\includegraphics[width=85mm]{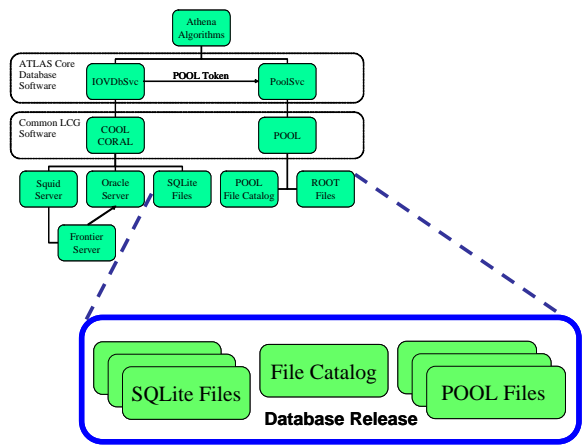}
\caption{Database Release technology hides the complexity of Conditions DB access (Fig.~\ref{db}).} \label{dbr}
\end{figure}

None of Tier-0 solutions for scalable database access is available on the Grid. To overcome scalability limitations of distributed 
database access~\cite{4}, we use the Database Release technology for deployment of the Conditions DB data on the Grid. Similarly to 
ATLAS software release packaging for distribution on the Grid, the Database Release integrates all necessary data in a single tar file:
\begin{itemize}\addtolength{\itemsep}{-0.5\baselineskip}
  \item the Geometry DB snapshot as an SQLite file,
  \item selected Conditions DB data as an SQLite file,
  \item corresponding Conditions DB POOL files and their POOL File Catalogue (Figure~\ref{dbr}).
\end{itemize}
Years of experience resulted in continuous improvements in the Database Release technology, which is used for 
ATLAS Monte Carlo simulations on the Grid. In 2007 the Database Release technology was proposed as a backup for database access in 
reprocessing at Tier-1 sites. 

\subsection{Challenges in Conditions DB Access}

In addition to Database Releases, Conditions DB data are delivered to all ten Tier-1 sites via continuous updates using Oracle 
Streams technology~\cite{5}. To assure scalable database access during reprocessing we stress-tested Oracle servers at the 
Tier-1 sites. As a result of stress-tests, we realized that the original model, where reprocessing jobs would run only at Tier-1 
sites and access directly their Oracle servers, causes unnecessary restrictions to the reprocessing throughput and most likely overload all Oracle servers when many jobs start at once.

In the first reprocessing campaign, the main problem with Oracle overload was exacerbated by additional scalability challenges.
Frirst, the reprocessing jobs for the cosmics data are five time faser than the baseline jobs reconstructing the LHC collision data, resulting in a fivefold increase in the Oracle load.
Second, having data on Tier-1s disks increases Oracle load sixfold (in contrast with the original model of reprocessing data from tapes). Combined with other limitations, these factors required increase in scalability by orders of magnitude.
To overcome the Conditions DB scalability challenges in reprocessing on the Grid,
the Database Release technology, originally developed as a backup, 
was selected as a baseline. 

\subsection{Conditions DB Release}

To overcome scalability limitations in Oracle access on the Grid, the following strategic decisions were made:
\begin{itemize}\addtolength{\itemsep}{-0.5\baselineskip}
%\begin{itemize}
  \item read most of database-resident data from SQLite,
  \item optimize SQLite access and reduce volume of SQLite replicas,
  \item maintain access to Oracle (to assure a working backup technology, when required).
\end{itemize}
As a result of these decisions, the Conditions DB Release technology fully satisfies reprocessing requirements, which has been proven on a scale of one billion database queries during two reprocessing campaigns of 0.5 PB of single-beam and cosmics data on the Grid~\cite{3}.
By enabling reprocessing at the Tier-2 sites, the Conditions DB Release technology effectively doubled CPU capacities at the BNL Tier-1 site during the first ATLAS reprocessing campaign. 

Conditions DB Release optimization for the second reprocessing campaign eliminated bottlenecks experienced earlier at few 
%\linebreak[4] 
Tier-1 sites with limited local network capabilities.
This Conditions DB Release was also used in user analysis of the reprocessed data on the Grid and during a successful world-wide 
LCG exercise called STEP'09.
In a recent fast reprocessing campaign, the Conditions DB Release integrated in a 1~GB dataset a slice of the Conditions DB data from two-weeks of data taking during this summer. The dataset was ``frozen'' to guarantee reproducibility of the reprocessing results.
During the latest ESD reprocessing campaign, further optimizations fit in a 1.4~GB volume a slice of Conditions DB for the data taking period of $0.23 \cdot 10^7$~s, which is about one quarter of the nominal LHC year. 

\begin{figure*}[t]
\centering
\includegraphics[width=130mm]{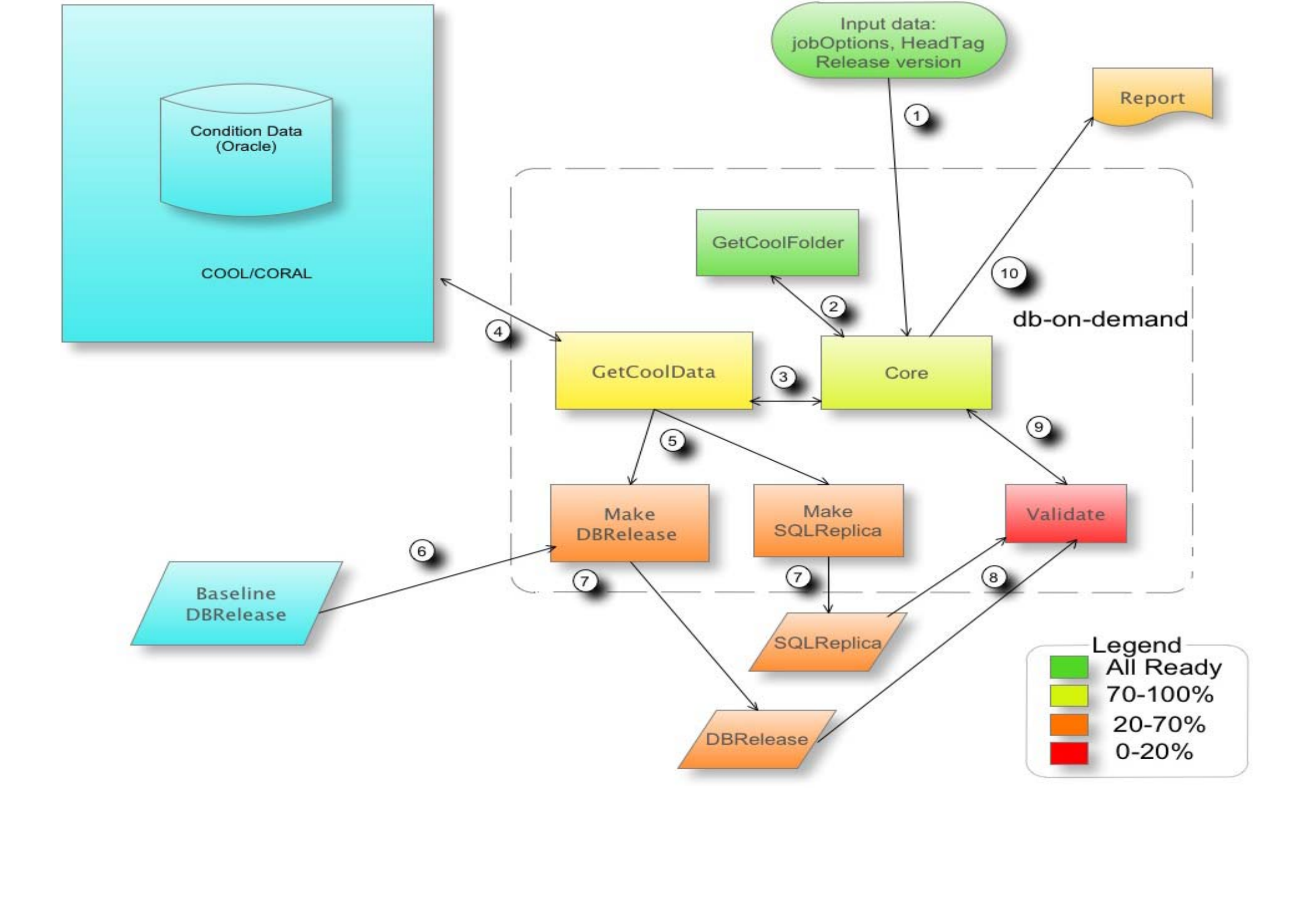}
\caption{Architecture of {\it db-on-demand} components automating Conditions DB Release build.} \label{demand}
\end{figure*}

To automate Conditions DB Release build sequence, we are developing the {\it db-on-demand} services
%(Section 5.1).
(Figure~\ref{demand}).
Recently these services were extended to support new requirements of the fast and ESD reprocessing that included check for missing interval-of-validity metadata.

\subsection{Direct Oracle Access}

For years ATLAS Monte Carlo simulations jobs used SQLite replicas for access to simulated Conditions DB data. Recently Monte Carlo simulations are becoming more realistic by using access to real Conditions DB data. This new type of simulation jobs requires access to Oracle servers. 
More realistic simulations provided an important new use case that validates our software for database access in a production 
environment. First realistic simulations used the software that has not yet been fully optimized for direct Oracle access. Thus the 
experience collected during summer was mixed: finished jobs peaked above 5000 per day; however, during remote database access some 
jobs used 1 min of CPU per hour, and others had transient segmentation faults and required several attempts to finish. 
There is a room for significant performance improvements with the software optimized for 
direct Oracle access~\cite{2}.

\begin{figure}[!t]
\centering
\includegraphics[width=90mm]{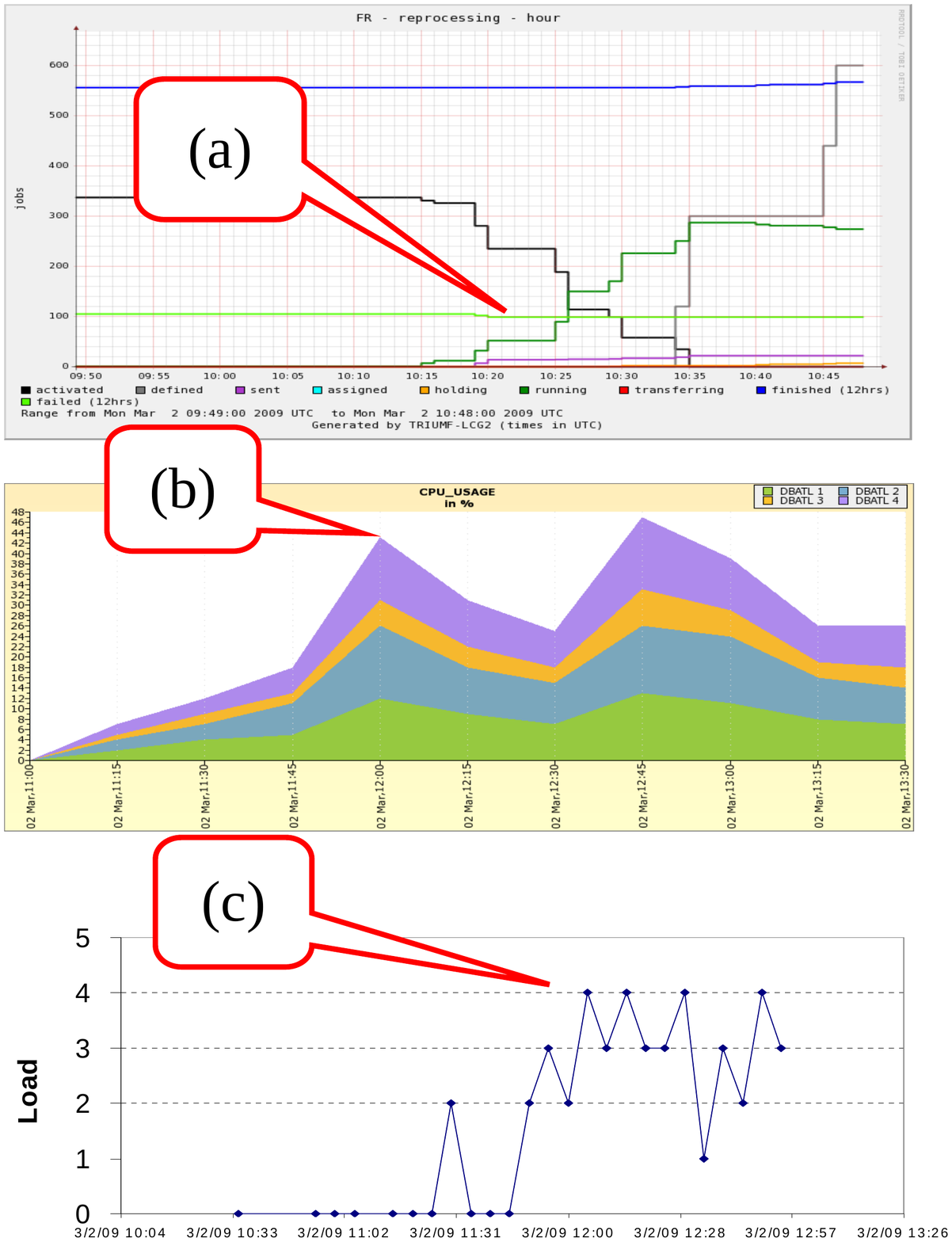}
\caption{Throttling Oracle server load on the Grid: (a) first batch of 300 jobs submitted;
(b) monitoring shows Oracle load is limited by the Pilot Query technology 
as we set ATLAS application-specific Oracle load limit at 4 (c).
} \label{pilot}
\end{figure}

To prevent bottlenecks in direct Oracle access in a Grid computing environment, we are developing a Pilot Query system for throttling job submission on the Grid.
Figure~\ref{pilot} shows the proof-of-principle demonstration of the Pilot Query approach at the Tier-1 site in Lyon. 
Development of the next generation Pilot Query system is now complete and ready for testing.

\section{Database Access Strategy}

Because Conditions DB access is crucial for operations with LHC data, we are developing the system where a different technology can 
be used as a redundant backup, in case of problems with a baseline technology.
While direct access to Oracle databases gives in theory the most flexible system, it is better to use the technology that is best 
suited to each use case~\cite{6}:
\begin{itemize}\addtolength{\itemsep}{-0.5\baselineskip}
%\begin{itemize}
  \item {\bf Monte Carlo simulations:} continue using the DB Release;
  \item {\bf first-pass processing:} continue using direct Oracle access at CERN;
  \item {\bf reprocessing:} continue using the Conditions DB Release;
  \item {\bf user analysis:}
\begin{itemize}\addtolength{\itemsep}{-0.5\baselineskip}
  \item {\it Grid jobs with large conditions data need:} use the Frontier/Squid servers;
  \item {\it local jobs with stable conditions data:} use the Conditions DB Release.
\end{itemize}
\end{itemize}
Status of late-coming components for database access in user analysis is described below.

\subsection{db-on-demand}

In user analysis, automated {\it db-on-demand} services eliminate the need for a central bookkeeping of database releases, since these will be created ``on-demand'' (Figure~\ref{demand}).
In order to have a user-friendly system, we will develop a web interface with user authentication based on secure technology for database access~\cite{7}, where each user would submit the request for a Conditions DB Release including all data needed to analyse a given set of events.

\subsection{DoubleCheck}

Frontier is a system for access to database-resident data via http protocol used by the CDF and CMS experiments~\cite{8}. To achieve 
scalability, the system deploys multiple layers of hardware and software
between a database server and a client: 
the Frontier Java servlet running within a Tomcat servlet container and the Squid---a single-threaded http 
proxy/caching server. In 2006 ATLAS tests done in collaboration with LCG found that Frontier does not maintain Squid cache consistency, which does not guarantee that ATLAS jobs obtain reproducible results in case of continuous updates to Conditions DB. In 2008 ATLAS resumed Frontier development and testing following recent breakthrough in addressing the Frontier cache consistency problem~\cite{9}.

In CMS case the cache consistency solution works for queries to a single table at a time. This does not work for ATLAS, as most our 
queries are for two tables. Hence the name DoubleCheck is chosen for a solution to the cache consistency problem developed for ATLAS. 
A major milestone in DoubleCheck development was achieved in July---the proof-of-principle test demonstrated that the LCG cache 
consistency solution developed for CMS can be extended to work for ATLAS. Further tests validated DoubleCheck for our major use 
case---updates of Conditions DB tables with the interval-of-validity metadata. DoubleCheck guarantees Frontier cache consistency within 15 minutes, which is close to delays observed in data propagation via Oracle Streams.

With no showstoppers in sight, ATLAS is now developing a plan and schedule for deployment, validation, and stress-testing of Frontier/Squid for database access in user analysis on the Grid.

%%%%%%%%%%%%%%%%%%%%%%%%%%%%%%%%%%

\section{Conclusions}

ATLAS has a well-defined strategy for redundant deployment of critical database-resident data. 
For each use case the most suited technology is chosen as a baseline:
\begin{itemize}\addtolength{\itemsep}{-0.5\baselineskip}
  \item Oracle for the first-pass processing at Tier-0; 
  \item Database Release for simulations and reprocessing on the Grid;
  \item Frontier for user analysis on the Grid.
\end{itemize}
The redundancy assures that an alternative technology can be used when necessary. 

ATLAS experience demonstrated that this strategy worked well as new unanticipated requirements emerged.
For example, the Conditions DB Release technology, originally developed as a backup, was choosen as a 
baseline to assure scalability of database 
access on the Grid. The baseline thechnology fully satisfies the requirements of several reprocessing procedures developed by the 
ATLAS collaboration. Steps are being taken to assure that Oracle can be used as a backup in case of unexpected problems with the 
baseline thechnology.

Each major ATLAS use case is functionally covered by more than one of the available technologies, so that we can achieve a redundant and robust data access system, ready for the challenge of the first impact with LHC collision data.

\bigskip % extra skip inserted
%%%%%%%%%%%%%%%%%%%%%%%%%%%%%%%%%%
\begin{acknowledgments}

I wish to thank all my collaborators who contributed to ATLAS database operations activities.

This work is supported in part by the U.S. Department of Energy, Division of High Energy Physics, under Contract DE-AC02-06CH11357.

\end{acknowledgments}

\medskip % extra skip inserted

\end{document}